\documentclass[9pt,twocolumn]{revtex4-1}
\usepackage{amsmath}
\usepackage{amssymb}
\usepackage{graphicx}
\usepackage{bm}
\usepackage{lineno}

\begin{document}

\title{ High-performance Raman quantum memory with optimal control}
\author{Jinxian Guo$^{1,2}$, Xiaotian Feng$^{1}$, Peiyu Yang$^1$, Zhifei Yu$^1$, L. Q. Chen$%
^{1*}$, Chun-Hua Yuan$^{1}$ and Weiping Zhang$^{2,3,*}$}

\affiliation{$^{1}$School of Physics and Material Science, East China Normal University, Shanghai 200062, P. R. China\\
$^{2}$Department of Physics and Astronomy, Shanghai Jiao Tong University, Shanghai 200240, China\\
$^{3}$Collaborative Innovation Center of Extreme Optics, Shanxi University, Taiyuan, Shanxi 030006, China\\
$^{*}$Corresponding authors lqchen@phy.ecnu.edu.cn, wpzhang@phy.ecnu.edu.cn}
\maketitle
\textbf{Quantum memories with high efficiency and
fidelity are essential for long-distance quantum communication and
information processing. Techniques have been developed for quantum memories based on atomic ensembles .
The atomic memories relying on the atom-light
resonant interaction usually suffer from the limitations of narrow
bandwidth. The far-off-resonant Raman process has been considered a potential
candidate for use in atomic memories with large bandwidths and high speeds. However,
to date, the low memory efficiency remains an unsolved bottleneck. Here, we
demonstrate a high-performance atomic Raman memory in }$^{87}$\textbf{Rb
vapour with the development of an optimal control technique. A memory
efficiency of 82.6$\%$ for 10-ns optical pulses is achieved and is the highest
realized to date in atomic Raman memories. In particular, an unconditional
fidelity of up to 98.0$\%$, significantly exceeding the no-cloning limit, is
obtained with the tomography reconstruction for a single-photon level
coherent input. Our work marks an important advance of atomic Raman memory
towards practical applications in quantum information processing.}

Quantum memory is a necessary component for quantum communications and
quantum computing. A practical quantum memory should be efficient, low-noise,
broadband and as simple as possible to operate \cite%
{fle,cold1,wal1,wal2,squ1,squ2}. Using several approaches, including
electromagnetically induced transparency (EIT), gradient echo memory (GEM),
the off-resonant Faraday effect and far off-resonant Raman memory, optical
memory has been demonstrated in cold atomic ensembles \cite{cold1,cold2},
atomic vapours \cite{wam2,Vapor2,Vapor3,Vapor4} and solids \cite%
{Solid1,Solid3, Solid2}. Yi-Hsin Chen et al. \cite{EIT1} reported a 96$\%$
memory efficiency for a coherent light pulse in a cold atomic ensemble using
EIT. M. Hosseini et al. \cite{PKLam} used GEM to realize a 78$\%$ memory
efficiency for weak coherent states with 98$\%$ fidelity. Polzik's group
\cite{Vapor3} demonstrated a quantum memory with a fidelity of 70$\%$ based
on the off-resonant Faraday effect. These examples \cite{EIT1,PKLam,Vapor3}
successfully demonstrated the capability to store optical states with high
efficiency and/or fidelity exceeding the classical limit \cite%
{efficiency1,efficiency3,He} and sub-megahertz bandwidths. However, the
bandwidth is important for the practical application of quantum memory \cite%
{Simon}. Quantum sources with bandwidth at the GHz level have been used in
long-distance quantum communication \cite{njp,qd} and quantum computers \cite%
{source}.

Unlike these protocols, far-off-resonant atomic Raman memory can store
short-time pulses corresponding to high bandwidths and can operate at high
speeds. In addition, the far-off-resonance characteristic makes the atomic Raman
memory \cite{shi,wam1,wam2} robust against inhomogeneities in the ensemble
and facilitates controlling the frequency of the output state. All of these
properties indicate that atomic Raman memory has great potential in
practical quantum information processing. The first experimental realization
of an atomic Raman memory was demonstrated \cite{wam1} in 2010. This
achievement represented significant progress in the field of Raman memory,
but some issues with low efficiency ($<$30$\%$) and significant noise from
the spontaneous four-wave mixing process persist. Recently, Raman memory
using photonic polarized entanglement \cite{shi} in a cold atomic ensemble
with a fidelity of 86.9$\pm $3.0$\%$, but an efficiency of only 20.9$\pm $7.7$\%$
was reported. An efficiency exceeding 50$\%$ and a fidelity exceeding
2/3 are necessary to store and retrieve an optical state within the no-cloning
regime without post-selection \cite{efficiency1,efficiency2,efficiency3,He}.
Therefore, previously, low efficiency appeared to exclude the broadband
Raman memory as an unconditional quantum memory.

In this paper, we present an optimal control technique where the atomic
vapour is performed a real-time optimal response on an input signal pulse.
With a $^{87}$Rb atomic vapour in paraffin-coated cell at T=78.5$^{o}$C, we
achieve a Raman quantum memory on a 10-ns-long coherent input of photon
number n$\approx $1 with 82.6\% memory efficiency and 98\% unconditional
fidelity.

\begin{figure}[tbh]
\includegraphics[width=3.0in]{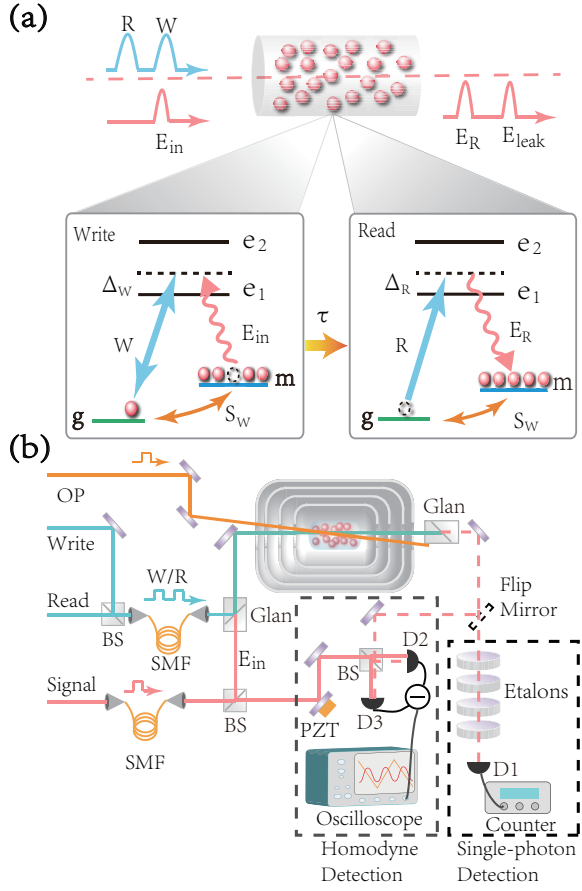}
\caption{\textbf{Raman memory}. (a) Schematic, atomic energy levels and
frequencies of the optical fields. $|g,m\rangle $: hyperfine levels $|$5$%
^{2} $S$_{1/2}$, F=1, 2$\rangle ;|e_{1}\rangle $ and $|e_{2}\rangle $:
excited states $|$5$^{2}$P$_{1/2}$, F=2$\rangle $ and $|$5$^{2}$P$%
_{3/2}\rangle $. $W $: write field; $E_{in}$: input signal; $E_{leak}$:
leaked signal; $S_{W}$; collective atomic spin wave; R: read field; $E_{R}$:
retrieved signal. (b) Experimental setup. The polarizations of the weak
signal beams, $E_{in}$ and $E_{R}$, are perpendicular to the strong driving
beams, W and R. The signals can be detected by a single-photon detector or
by homodyne detection. OP: optical pumping laser; SMF: single-mode fibre;
D1: single-photon detector; D2, D3: photo-detector; BS: beam splitter; PZT:
piezoelectric transducer.}
\label{Fig1}
\end{figure}

\section{Experimental setup}

The experimental setup and atomic levels are depicted in Fig. 1. The $^{87}$%
Rb atomic vapour in the paraffin-coated glass cell is the core component of
the current Raman memory. The atomic cell is 10.0 cm long, has a diameter of 1.0
cm and is heated to 78.5$^{o}C$. Our Raman memory starts with a large
ensemble of atoms that were initially prepared in the $|m\rangle $ state by a 44-$\mu $%
s-long OP pulse. Then, the input signal pulse $E_{in}$ is stored as atomic
spin excitation $S_{W}$ induced by the strong off-resonant write pulse ($W$)
with the Rabi frequency $\Omega _{W}(t)$ and detuning $\Delta _{W}$. After a
certain delay $\tau $, the atomic excitation can be retrieved into optical
state $E_{R}$ by the strong off-resonant read pulse ($R$) with the Rabi
frequency $\Omega _{R}(t)$ and detuning $\Delta _{R}$. The waists of the
laser beams ($W$, $R$ and $E_{in}$) are all 600 $\mu m$. The two strong
driving beams, $W$ and $R$, can be generated by the same or different lasers
and are coupled into the same single-mode fiber. Their intensities and
temporal shapes are controlled by acousto-optic modulators (AOMs). The input
$E_{in}$ signal comes from another laser phase-locked on the $W$ laser. The
temporal shape is controlled by a Pockels cell (Conoptics, model No.
360-80). The shortest pulse duration of the Pockels cell is 10 ns. The $W$
and $E_{in}$ fields are two-photon resonant and spatially overlapped after
passing through a Glan polarizer with 94\% spatial visibility in the atomic
vapour. The output signals can be separated from the strong driving pulses
by another Glan polarizer with an extinction ratio of 40 dB and are detected by
a homodyne detection or by a single-photon detector after passing through four
etalons to filter the leaked driving photons at 115 dB with 33$\%$
transmission of the signal photons.

\begin{figure*}[tbh]
\includegraphics[width=6.0in]{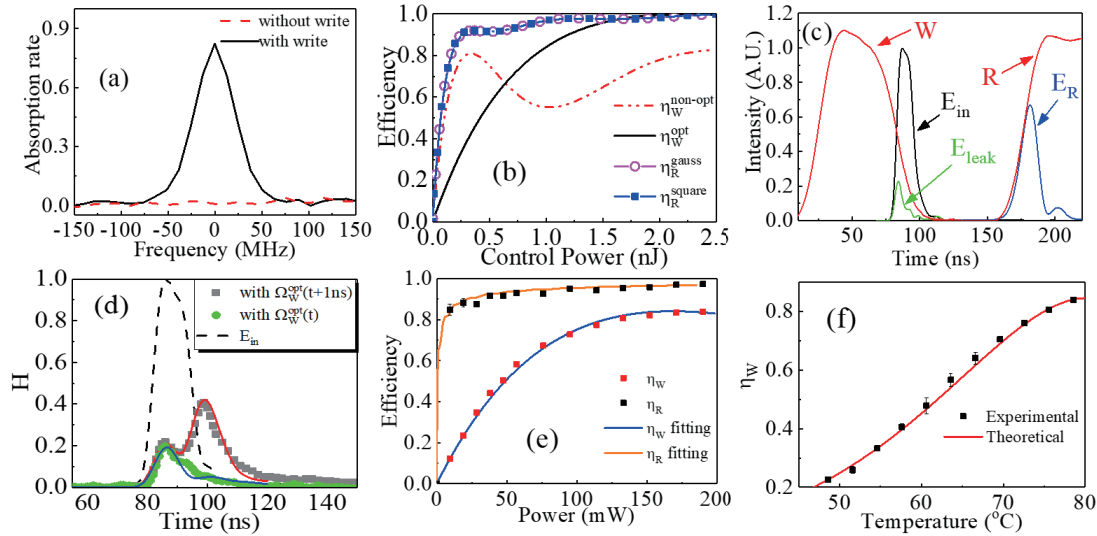}
\caption{Efficient Raman memory. (a) Absorption rate of the weak
input-signal pulse as a function of the Raman detuning frequency. $\Delta
_{W}$ is fixed at 3.0 GHz. (b) Theoretical efficiency as a function of
the energy of the strong control pulse. The input optical pulse is a 10-ns
near-square pulse. All optical fields detune 3.0 GHz from atomic transition
and the optical depth $d=1100$. In the write process, the efficiency is always
much smaller than 1.0 when using a non-optimal write pulse (10-ns Gaussian shape),
but it can approach 1.0 with the optimal write pulse when the write pulse
is larger than 1.5 nJ. In the read process, the curves with Gaussian and square
read pulses coincide with each other. The retrieval efficiency is waveform-independent
and increases with the energy of the read pulse until approaching
1.0. (c) Temporal modes of the strong driving (red, $W$, $R$), input
signal (black, $E_{in}$), leaked signal (green, $E_{leak}$) and output
signal (blue, $E_{R}$) pulses. (d) Waveform of the leaked signal with
the $\Omega _{W}^{opt}(t)$ (green dot) and $\Omega _{W}^{opt}(t+1ns)$ (gray
square) write pulse. The lines are the corresponding theoretical fits.
(e) Storage ($\protect\eta _{W}$) and retrieval ($\protect\eta _{R}$)
efficiencies as functions of the power of the driving field with the
optimized write pulse and the corresponding theoretical fitting. (f) The
write efficiency as a function of the cell temperature.}
\label{Fig2}
\end{figure*}

\section{Experimental Results}

\textbf{Efficiency}

The Raman write process is a type of coherent absorption induced by a strong
write pulse. As shown in Fig. 2(a), when the write pulse is switched off, owing
to the far-off-resonant frequency, almost 100\% of the $E_{in}$ pulse passes
through the atomic vapour. Below, we use the total energy of such an $E_{in}$
pulse to normalize the write and retrieve efficiencies. When the write pulse
is turned on, part of the energy of the $E_{in}$ pulse is converted
coherently as the atomic spin wave $S_{W}(z)$ near the two-photon resonance
frequency. The rest of the  $E_{in}$ energy passes through the atoms as $E_{leak}$,
as shown in Fig. 1(a). The full width at half maximum (FWHM) of the absorption
spectrum is approximately 50~MHz, as shown in Fig. 2(a).

According to the theoretical analysis in Ref. \cite{Gorsh}, the
spatial-distributed atomic spin wave in a far-off-resonant Raman write process
is given by
\begin{equation}
S_{W}(z)=\int_{0}^{t_{W}}dtq(z,t)E_{in}(t)  \label{Eq1}
\end{equation}%
where $q(z,t)=i\frac{\sqrt{d}}{\Delta _{W}}\Omega _{W}^{\ast }(t)e^{i\frac{%
dz+h(t,t_{W})}{\Delta _{W}}}J_{0}(\frac{\sqrt{4h(t,t_{W})dz}}{\Delta _{W}^{2}})$, $%
t_{W}$ is the duration of the write process, $d$ is the optical depth of
atomic ensemble and $h(t,t_{W})=\int_{t}^{t_{W}}\left\vert \Omega (t^{\prime
})\right\vert ^{2}dt^{\prime }$. Eq. (1) is an iterative function that
is determined by the matching between the temporal shapes of the input $%
E_{in}(t)$ and the write pulse $\Omega _{W}(t)$ \cite{svd1,svd2,Gorsh}. Therefore,
to achieve efficient conversion, it is crucial to perform real-time
control on $\Omega _{W}(t)$ or $E_{in}(t)$\ to make the atoms coherently
absorb as much energy $E_{in}(t)$ as possible. The optimal control of $%
E_{in}(t)$ has been used to achieve efficient memory in an EIT-based process
\cite{EIT1}, where the shape of the input signal $E_{in}(t)$ was adjusted
according to atomic memory system. Here, we prefer the dynamical control $\Omega
_{W}(t)$ because a quantum memory system should have the ability to store
and preserve quantum information of an input optical signal with an arbitrary
pulse shape. To obtain the optimal $\Omega _{W}(t)$, denoted $\Omega
_{W}^{opt}(t)$, we first use the iterative methods mentioned in Ref. \cite%
{Gorsh} to calculate the optimal spin wave, corresponding to the minimum
 $E_{leak}$. Then, the optimal spin wave establishes a one to one
correspondence between $E_{in}(t)$ and $\Omega _{W}^{opt}(t)$ via Eq. (\ref%
{Eq1}). Thus, for any given shape of $E_{in}(t)$, $\Omega
_{W}^{opt}(t)$ can be obtained from Eq. (\ref{Eq1}) via the optimal spin
wave. Moreover, the corresponding optimal efficiency $\eta _{W}^{opt}$
depends on only the optical depth $d$ and the total energy of the write pulse.
Fig. \ref{Fig2}(b) shows the theoretical efficiencies as the function of
the energy of the strong driven pulses. Using a 10-ns near-square pulse as
the input $E_{in}$, the write efficiency with $\Omega _{W}^{opt}(t)$ is
approximately equal to 1 when the write energy is larger than 1.5 nJ, while
the maximum write-in efficiency with a non-optimized $\Omega _{W}(t)$ (a
10-ns Gaussian-shaped $\Omega _{W}(t)$ is used in Fig. \ref{Fig2}) is much
smaller than one. In the read process (see Fig \ref{Fig1}), the spin wave $%
S_{W}(z)$ is retrieved back to the optical field $E_{R}(t)$ by the read pulse $%
\Omega _{R}(t)$. Unlike $\eta _{W}$, the retrieval efficiency $\eta _{R}$ is
independent of the temporal waveform \cite{Gorsh}, and a read pulse with strong
power but without temporal optimization is sufficient for $\eta _{R}\sim 1$.
This can be seen in Fig. \ref{Fig2}(b). $\eta _{R}$ increases with the
total energy of the read pulse, whether Gaussian or square-shaped until $%
\eta _{R}\sim 1$. Thus, with the above optimal control on $\Omega _{W}(t)$, the
total efficiency of the Raman memory process is $\eta _{T}=\eta _{W}\times \eta
_{R}\sim 1$ in principle.

Here, we experimentally demonstrate a break of the efficiency in Raman
memory with dynamic control over the temporal shape of the write pulse. In
the experiment, the given $E_{in}$ pulse is mapped in a forward-retrieval
configuration. We derive $\Omega _{W}^{opt}(t)$ using the iteration-based
optimization strategy based on the given short $E_{in}(t)$ pulse and
experimentally control the temporal profile of the write pulse by using an
intensity modulator (here, an AOM). The temporal profile of the optimized
write pulse is shown in Fig. 2(c). To show the magic improvement of
optimization control, two write pulses are given, the optimal $\Omega
_{W}^{opt}(t)$ and an optimal write pulse delayed by 1.0 ns, $\Omega
_{W}^{opt}(t$+1ns$)$. The corresponding leaked optical pulses and the
theoretical fits are shown in Fig. 2(d). The leaked energy for the square
curve is twice that for the dot curve. Through the optimal control, the leaked energy
of the input signal is greatly reduced. The storage efficiency $\eta _{W}$,
calculated by $(\overline{N}_{E_{in}}$-$\overline{N}_{E_{leak}})/\overline{N}%
_{E_{in}}$, reaches $\sim $84$\%$ when the atomic temperature $T$ is 78.5$%
^{o}$C and the power of the write pulse is 190 mW (Fig. 2(e)). The
retrieval efficiency $\eta _{R}$, calculated by $(\overline{N}_{E_{R}}$/($%
\overline{N}_{E_{in}}$-$\overline{N}_{E_{leak}})$, can reach 98.5$\%$
when the read laser is 190 mW, with 3.0 GHz frequency detuning (Fig. 2 (e)).
The total memory efficiency, $\eta _{T}$=$\eta _{W}\times \eta _{R}$, is 82.6%
$\%$ when the input signal pulse contains an average number of photons
ranging from 0.4 to 10$^{4}$; thus, this Raman memory is a good linear
absorber. The 82.6$\%$ memory efficiency is the best performance reported to
date for Raman-based memory and far exceeds the no-cloning limit.

\begin{figure*}[tbh]
\includegraphics[scale=0.55]{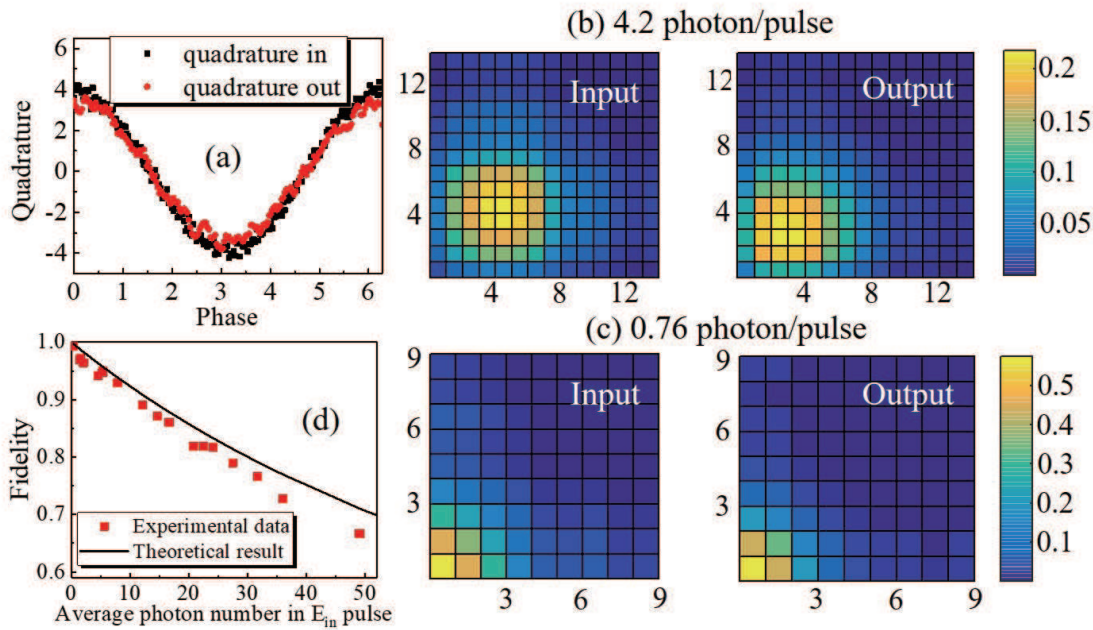}
\caption{\textbf{Fidelity of the Raman memory}. (a) Quadrature amplitudes of
the input and output signal pulses at an average of 7.9 photons/pulse. The
density matrices of the input and output signal pulses at 4.2 (b) and 0.76
(c) photons/pulse on average. (d) Fidelity as a function of the number of
photons contained in the input signal pulse. The red squares show the
experimental data, and the black line shows the theoretical result. }
\label{Fig3}
\end{figure*}

\textbf{Fidelity}

Fidelity is the ultimate performance criterion for quantum memory and
reflects the maintenance of the quantum characteristics of the optical
signal during the memory process. At the few-photon level, fidelity is
readily degraded by excess noise and is mainly caused by the four-wave mixing
(FWM) process \cite{wam2} and spontaneous emission. Spontaneous noise
comes from the spontaneous Raman scattering between the strong write pulse and
the atoms populating the $|g\rangle $ state. Having fewer $|g\rangle $ atoms helps
suppress the spontaneous excess noise. In our paraffin-coated cell, more
than $98\%$ of the atoms populate the $|m\rangle $ state. The spontaneous
emission noise intensity is measured by determining the photon number using
a single-photon detector when the $E_{in}$ pulse is turned off. On average,
the spontaneous noise is approximately 0.02 photons per memory process
before the etalons for two strong driving pulses with a power of 190 mW at
a detuning frequency of 3.0 GHz. The FWM excess noise is mainly attributed
to anti-Stokes ($AS_{FWM}$, with same frequency of $E_{R}$) and Stokes ($%
S_{FWM}$) photons with the same intensity. We can deduce the proportion of $%
AS_{FWM}$\ in retrieved $E_{R}$\ pulse by measuring the intensity of $%
S_{FWM} $\ using single-photon detection. Our results show that the $%
AS_{FWM}$\ noise is less than 10\% in $E_{R}$. Such low excess noise
effectively guarantees the fidelity of the quantum memory process.

To achieve the fidelity performance of the current Raman quantum memory, we
measure the fidelity using the equation $F=\left\vert Tr\left( \sqrt{\sqrt{%
\rho _{in}}\rho _{out}\sqrt{\rho _{in}}}\right) \right\vert ^{2}$ \cite%
{Fidelity}, where $\rho _{in}$ and $\rho _{out}$ are the reconstructed
density matrices of $E_{in}$ and $E_{R}$, respectively. We record the
quadrature amplitudes of the $E_{in}$ and $E_{R}$ signals using homodyne
measurement, and we then reconstruct the density matrices by tomographic
reconstruction \cite{Homodyne}. The setup used for homodyne detection is
shown in Fig. 1(b). To stabilize the phase difference between the $E_{in}$
and $E_{R}$ pulses and simplify the homodyne setup, the write and read
pulses are generated by the same laser and are controlled using one AOM. In
the measurement, the two weak signals, $E_{in}$ and $E_{R}$, are both short
pulses. Matching the temporal modes of short pulses is difficult. Therefore,
we use a strong continuous laser beam with the same frequency as the signal
pulses $E_{in}$ and $E_{R}$ as the local oscillator for homodyne detection
(the detailed strategy can be found in Refs. \cite{Homo,Homodyne}). We
recorded 10$^{5}$ sets of quadrature amplitudes of the $E_{in}$ and $E_{R}$
pulses while varying the phase of the local oscillator between 0 and 2$\pi $
by scanning the piezoelectric transducer, multiplying the quadrature
amplitudes of each pulse by the pulse shapes of the corresponding signals,
and finally, integrating the product over the signal pulse duration. The
integrated quadrature amplitude as a function of the local oscillator phase
is shown in Fig. 3(a), where the mean number of photons contained in the $%
E_{in}$ pulse is 7.9. The phase of the retrieved $E_{R}$ signal pulse
closely follows that of the input $E_{in}$ pulse.

The density matrix elements of the $E_{in}$ and $E_{R}$ pulses are obtained
based on the quadrature-amplitude results using the maximum-likelihood
reconstruction method \cite{ML,Homodyne}. The results are plotted in Fig.
3(b,c), with the input pulses containing, on average, 4.2 and 0.76 photons,
corresponding to unconditional fidelities of 0.915 and 0.98, respectively.
The fidelities significantly exceed the no-cloning limit, indicating that
the current Raman memory is a quantum-memory process and does not introduce
significant excess noise during the memory process.

As mentioned above, the current Raman memory is a good linear absorber and
allows the storage and retrieval of coherent optical signals at the
single-photon level for up to 10$^{4}$ photons with the same memory
efficiency. Unlike the efficiency, the unconditional fidelity of the quantum
memory of the coherent field is related to the average photon number contained
in the input signal ($\bar{N}_{E_{in}}$) and efficiency ($\eta _{T}$) by
 $F=1/[1+\bar{N}_{E_{in}}(1-\sqrt{\eta _{T}})^{2}]$\textbf{\ }%
\cite{He}\textbf{,}  which shows that if $\eta _{T}$ $<$1, the fidelity will
rapidly decrease with $\bar{N}_{E_{in}}$ owing to the worse overlap between $%
\rho _{in}$\ and $\rho _{out}$. In Fig. 3(d), the fidelity is shown as a
function of $\bar{N}_{E_{in}}$ with $\eta _{T}$=82.6$\%$. The experimental $%
F $ value is slightly smaller than the theoretical $F$ value because of the
excess noise in the experiment. $F$\textbf{\ }exceeds the no-cloning limit \cite%
{LamE,Noclon1,Noclon2} at $\bar{N}_{E_{in}}\leq $49 in the current Raman memory
process. \textbf{\ }

\textbf{Bandwidth and coherence time}

Raman memory is a genuine broadband memory. The ability to store and
retrieve broadband pulses was successfully demonstrated in Ref. \cite{wam1},
where a bandwidth larger than 1 GHz of the retrieved signal was obtained using
a 300 ps and 4.8 nJ read pulse. In a practical Raman memory, the bandwidth
is generated dynamically by the strong driving pulses. In Fig. 2(c), the $%
E_{in}$ pulse  has a FWHM of 10 ns and a bandwidth of 100 MHz. The FWHM of
the $E_{R}$ signal pulse is 13 ns, corresponding to a bandwidth of 77 MHz.
The 77 MHz bandwidth of the single $E_{R}$ signal in the current memory is
dozens or hundreds of times larger than the values reported based on the EIT,
Faraday and GEM approaches, thus demonstrating the broadband memory ability of
the current quantum memory scheme.

The decoherence time, which is another essential criterion for good quantum
memory, is measured to be approximately 1.1 $\mu $s in the present atomic
system and is mainly limited by the diffusion of the atoms \cite{defuse}. The
delay-bandwidth product at 50$\%$ memory efficiency, an appropriate figure
of merit, is defined as the ratio of the memory time to the duration of the
signal pulse and is 52 in this work.

In summary, we have demonstrated a high-performance broadband quantum
optical memory via pulse-optimized Raman memory in free space. The 82.6$\%$
memory efficiency is the highest value obtained to date for far-off-resonant
Raman memory. The unconditional fidelity of 98$\%$ for an input pulse
containing an average of approximately one photon significantly exceeds the
classical limit. The 77 MHz bandwidth of the current memory is dozens or
hundreds of times larger than the reported bandwidths for memories based on the EIT,
Faraday and GEM approaches. The delay-bandwidth product at 50$\%$ memory
efficiency is 52. These attractive properties demonstrate that the Raman
memory is a high-performance broadband quantum memory. Additionally, our
memory is implemented in an atomic vapour system that can be easily operated
and could become the core of a scalable platform for quantum
information processing, long-distance quantum communication and quantum
computation.

Further improvement of the performance could be achieved by improving the
experimental conditions. Higher efficiency could be achieved by increasing
the atomic optical depth through increasing the atomic temperature (Fig. 2(e)),
lengthening the cell, or obtaining better spatial-mode matching. The
bandwidth could be improved to 1 GHz by using shorter and stronger driving
pulses\cite{wam1}. The decoherence time or the delay-bandwidth product at 50$%
\%$ memory efficiency is limited by atomic diffusion out of the laser
beam in the current memory system. It could be increased to as high as 10$%
^{5}$ by using an anti-relaxation-coated cell \cite{Vapor3} with the same
diameter as the laser beams, for which the typical decoherence time is approximately
several milliseconds.

\begin{center}
\textbf{Acknowledgements}
\end{center}

This work was supported by the National Key Research and Development Program
of China under grant number 2016YFA0302001, and by the National Natural
Science Foundation of China (grant numbers 91536114, 11474095, and
11654005), and National Science Foundation of Shanghai (No. 17ZR1442800).

\begin{center}
\textbf{Competing financial interests}
\end{center}

The authors declare no competing financial interests.

\end{document}